\documentclass[a4paper]{article}

\usepackage{amsmath}
\usepackage{amsfonts}
\usepackage{dsfont}

\newcommand{\m}[1]{$#1$}

\newcommand{\refeqp}[1]{(\ref{eq:#1})}

\newcommand{\paraa}[1]{\big(#1\big)}
\newcommand{\parab}[1]{\Big(#1\Big)}
\newcommand{\parac}[1]{\bigg(#1\bigg)}

\newcommand{\real}{\mathbb{R}}
\newcommand{\compl}{\mathbb{C}}

\newcommand{\half}{\frac{1}{2}}

\newcommand{\xv}{\vec{x}}

\renewcommand{\mid}{\mathds{1}}

\newcommand{\U}[1]{U_{#1}}
\newcommand{\Mv}{{\overrightarrow{\! M}}}
\newcommand{\Mb}{\mathbf{M}}
\newcommand{\mv}{\overrightarrow{m}}
\newcommand{\mpv}{{\overrightarrow{m}'}}
\newcommand{\nv}{\overrightarrow{n}}
\newcommand{\npv}{{\overrightarrow{n}'}}
\newcommand{\phiv}{\overrightarrow{\varphi}}
\newcommand{\A}{\mathcal{A}}
\newcommand{\su}{\textrm{su}}
\newcommand{\ea}{e_{\alpha}}
\newcommand{\ema}{e_{-\alpha}}
\newcommand{\eb}{e_{\beta}}
\newcommand{\emb}{e_{-\beta}}
\newcommand{\eab}{e_{\alpha+\beta}}
\newcommand{\emab}{e_{-\alpha-\beta}}
\newcommand{\X}{X}

\newcommand{\Xdd}{\overset{\, ..}{\X}}
\newcommand{\Xd}{\overset{\, .}{\X}}

\newcommand{\xd}{\overset{.}{x}}
\newcommand{\const}{\textrm{const}}
\newcommand{\CCom}[3]{\Big[\big[#1,#2\big],#3\Big]}
\newcommand{\Com}[2]{\big[#1,#2\big]}
\newcommand{\Comb}[2]{\Big[#1,#2\Big]}
\newcommand{\grs}{g_{rs}}
\newcommand{\gurs}{g^{rs}}
\newcommand{\dr}{\partial_{r}}
\newcommand{\ds}{\partial_{s}}
\newcommand{\done}{\partial_{1}}
\newcommand{\dtwo}{\partial_{2}}
\newcommand{\Us}{U}
\newcommand{\UmN}{\Us_{\mv}^{(N)}}
\newcommand{\UnN}{\Us_{\nv}^{(N)}}
\newcommand{\UmnN}{\Us_{\mv+\nv}^{(N)}}
\newcommand{\Um}{\Us_{\mv}}
\newcommand{\Umb}{\Us_{-\mv}}
\newcommand{\Ump}{\Us_{\mv'}}
\newcommand{\Umpb}{\Us_{-\mv'}}
\newcommand{\Un}{\Us_{\nv}}
\newcommand{\Unb}{\Us_{-\nv}}
\newcommand{\Unp}{\Us_{\nv'}}
\newcommand{\Unpb}{\Us_{-\nv'}}
\newcommand{\glNC}{\textrm{gl}(N,\compl)}
\newcommand{\glNhC}{\textrm{gl}(\Nh,\compl)}
\newcommand{\Nh}{\hat{N}}
\newcommand{\Nhp}{{(\hat{N})}}
\newcommand{\Np}{{(N)}}

\newcommand{\Ufactst}{\frac{3}{32\pi}}
\newcommand{\Ufactsti}{\frac{3}{32\pi i}}

\newcommand{\sqo}{\sqrt{\omega}}
\newcommand{\xpg}{\vec{x}_+^{[\gamma]}}
\newcommand{\xmg}{\vec{x}_-^{[\gamma]}}
\newcommand{\xp}{\vec{x}_+}
\newcommand{\xm}{\vec{x}_-}

\title{Lie Algebras and Quantum Groups \\ \large{Homework exercises set 8}}
\author{Joakim Arnlind, 790507-0657 \\ \small{joakim.arnlind@math.kth.se}}
\date{December, 2003}

\begin{document}

\thispagestyle{empty}

\vspace{2cm}

\begin{center}
\Large{More Membrane Matrix Model Solutions,\\ -- and Minimal
  Surfaces in \m{S^7}}\\
\vspace{0.25cm}
\large{Joakim Arnlind and Jens Hoppe}\\
\small{Department of Mathematics\\
Royal Institute of Technology\\
Stockholm\\
\vspace{0.3cm}
 December 2003}

\vspace{3cm}

\textbf{\large{Abstract}}\\
\end{center}


\noindent New solutions to the classical equations of motion of a
bosonic matrix-membrane are given. Their continuum limit defines
3-manifolds (in Minkowski space) whose mean curvature vanishes.
Part of the construction are minimal surfaces in \m{S^7}, and
their discrete analogues.

\newpage

\pagenumbering{arabic}

\noindent Some time ago \cite{Hoppe:1999}, solutions of the bosonic matrix-model
equations,
\begin{align}
\begin{split}
  &\Xdd_i = -\sum_{j=1}^d \CCom{X_i}{X_j}{X_j}\\
  & \sum_{i=1}^d [\X_i,\Xd_i] = 0\label{eq:commeqs}
\end{split}
\end{align}
were found where
\begin{align}
  \X_i(t) = x(t)\mathcal{R}_{ij}(t)M_j,\label{eq:XRM}
\end{align}
with \m{\mathcal{R}(t)=e^{\A\varphi(t)}} a real, orthogonal \m{d\times d}
matrix, \m{x(t)} and \m{\varphi(t)} being given via
\begin{align}
  \begin{split}
  \half\xd^2+\frac{\lambda}{4}x^4+\frac{L^2}{2x^2} &=\const. \\
  \varphi^2(t)\xd(t) &= L (=\const),\label{eq:consteq}
  \end{split}
\end{align}
and the \m{d} hermitean \m{N\times N} matrices \m{M_i} satisfying
\begin{align}
  \begin{split}
    \sum_{j=1}^{d'} \CCom{M_i}{M_j}{M_j} &= \lambda M_i\\
    i&=1,\ldots,d'.\label{eq:Mcom}
  \end{split}
\end{align}
The reason for \m{d'} (rather than \m{d}) appearing in \refeqp{Mcom}
was that in order to satisfy the two remaining conditions,
\begin{align}
  \A^2\Mv = -\Mv&\label{eq:Asq}\\
  \sum_{j=1}^d \Com{M_j}{(\A\Mv)_j} = 0&\label{eq:MAcom}
\end{align}
-- which have to be fulfilled in order for \refeqp{XRM} to satisfy
   \refeqp{commeqs} -- in an ``irreducible'' way (the matrix valued
   \m{d}-component vector \m{\Mv} can, of course, always be broken up
   to contain pairs of identical pieces) half -- or more -- of the
   matrices \m{M_j} were chosen to be zero, and (permuting the \m{M}'s
   such that the first \m{d'\leq\frac{d}{2}} are the non-zero ones)
   the non-zero elements of \m{\A} as \m{\A_{i+d',j} = 1 =
   -\A_{j,i+d'}}, \m{i,j=1,\ldots,d'};
in particular, \refeqp{MAcom} was satisfied by having, for each
   \m{j}, either \m{M_j} or \m{(\A\Mv)_j} be identically zero.

As, in the membrane context. \m{d\overset{(<)}{=}9}, \m{d'=4} recieved
particular attention, while the continuum
limit of \refeqp{Mcom},
\begin{gather}
  \sum_j \{\{m_i,m_j\},m_j\} = -\lambda m_i,\label{eq:mpbpb}\\
  \parac{
    \{m_i,m_j\}:=\frac{1}{\rho}\paraa{\partial_1 m_i\partial_2 m_j -
  \partial_2 m_i \partial_1 m_j};\,
    \grs:= \dr\mv\cdot\ds\mv; \,\mv=\mv(\varphi^1,\varphi^2)
  }\notag
\end{gather}
alias
\begin{align}
  \frac{1}{\rho}\dr\frac{g\gurs}{\rho}\ds\mv = -\lambda\mv\label{eq:rholaplace}
\end{align}
is related to
\begin{align}
  \begin{split}
  \frac{1}{\sqrt{g}}\dr\sqrt{g}\gurs\ds\mv = -2\mv,& \\
  \mv^2=1,&
  \end{split}\label{eq:laplace}
\end{align}
i.e the problem of finding minimal surfaces in higher dimensional
spheres (which for \m{d'=4} was proven \cite{Lawson} to admit solutions of any genus).

In this letter, we would like to enlarge the realm of explicit
solutions (of \refeqp{commeqs}, resp. its \m{N\rightarrow\infty}
limit, resp \refeqp{laplace}) while shifting emphasis from \m{d'=4} to
\m{d'=8}
(the case \m{d'=6}, which can be used to obtain nontrivial solutions
in the BMN matrix-model, will be discussed elsewhere).

Our first observation is that \refeqp{MAcom} rather naturally admits
solutions which avoid the ``doubling mechanism''. While \m{\A} is kept
to be an ``antisymmetric permutation''-matrix in a maximal
even-dimensional space, \refeqp{MAcom} can be realized if
\m{\Mb:=\{M_j\}_{j=1}^d} (with \m{M_d\equiv 0} if \m{d} is
odd) can be written as a union of even-dimensional subsets of mutually
commuting members.
In order to give a first example, let us, for later convenience, define
(for arbitrary odd \m{N>1}) \m{N^2} independent \m{N\times N} matrices
\begin{align}
  \UmN:= \frac{N}{4\pi M(N)}\omega^{\half m_1 m_2} g^{m_1} h^{m_2}
  \label{eq:Udef}
\end{align}
where \m{\omega:=e^{\frac{4\pi iM(N)}{N}}}, \m{\mv=(m_1,m_2)},
\begin{align}
  g =
  \begin{pmatrix}
    1 & 0 & \cdots & 0\\
    0 & \omega & \ddots & \vdots\\
    \vdots & \ddots & \ddots & 0\\
    0& \cdots & 0 & \omega^{N-1}
  \end{pmatrix},\quad
  h=
  \begin{pmatrix}
    0 & 1 & 0 & \cdots & 0 \\
    0 & 0 & 1 & \cdots & 0 \\
    \vdots & \vdots &\ddots & \ddots & \vdots\\
    0 & 0 & \cdots & 0 & 1 \\
    1 & 0 & \cdots & 0 & 0
  \end{pmatrix}.\label{eq:ghomega}
\end{align}
\refeqp{Udef} provides a basis of the Lie-algebra \m{\glNC}, with
\begin{align}
  \Comb{\UmN}{\UnN} = -\frac{iN}{2\pi M(N)}
  \sin\parac{\frac{2\pi M(N)}{N}\paraa{\mv\times\nv}}\UmnN\label{eq:Ucom}
\end{align}
(for the moment, we will put \m{M(N)=1}, as only when
\m{N\rightarrow\infty}, \m{\frac{M(N)}{N}\rightarrow \Lambda\in\real},
this ``degree of freedom'' is relevant).

Let now \m{N=3},
\begin{align}
\begin{split}
  \Mv &= \half\parac{
    \frac{\U{1,0}+\U{-1,0}}{2},
    \frac{\U{1,0}-\U{-1,0}}{2i},
    \frac{\U{0,1}+\U{0,-1}}{2},
    \frac{\U{0,1}-\U{0,-1}}{2i},
    \frac{\U{1,1}+\U{-1,-1}}{2},\\
    &\frac{\U{1,1}-\U{-1,-1}}{2i},
    \frac{\U{-1,1}+\U{1,-1}}{2},
    \frac{\U{-1,1}-\U{1,-1}}{2i}
    }\\
  &=:\paraa{M_1,M_2,M_3,M_4,M_5,M_6,M_7,M_8}.
\end{split}\label{eq:Msu}
\end{align}
\refeqp{Msu} satisfies \refeqp{Mcom}, \m{[M_1,M_2] = 0,[M_3,M_4] = 0,[M_5,M_6]
  = 0} and \m{[M_7,M_8] = 0}
(note that we have implicitly reordered the elements of \m{\A}), and \m{\Mv^2=\mid}.

One can rewrite the 8 \m{M_j}'s, being a basis of \m{\su(3)}, in terms of
the Cartan-Weyl basis \m{\{h_1,h_2,\ea,\ema,\eb,\emb,\eab,\emab\}},
\begin{align}
  \begin{split}
    &[h_1,h_2]=0\\
    &[h_i,\ea]=\alpha_i\ea\quad [h_i,\ema]=-\alpha_i\ema
    \quad\alpha = (2,0)\\
    &[h_i,\eb]=\beta_i\eb\quad [h_i,\emb]=-\beta_i\emb
    \quad\beta = (-1,\sqrt{3})\\
    &[h_i,\eab]=(\alpha+\beta)_i\eab\quad
    [h_i,\emab]=-(\alpha+\beta)_i\emab\\
    &[\ea,\ema]=4h_1\qquad [\eb,\emb]=-2h_1+2\sqrt{3}h_2\\
    &[\eab,\emab] = 2h_1+2\sqrt{3}h_2\\
    &[\ea,\eb]=2\eab\qquad [\ea,\emab]=-2\emb\\
    &[\ema,\eab]=2\eb\qquad [\ema,\emb]=-2\emab\\
    &[\eb,\emab]=2\ema\qquad [\emb,\eab]=-2\ea,
  \end{split}\label{eq:CWbasis}
\end{align}
obtaining
\begin{align}
\begin{split}
  M_1 &= \Ufactst  \paraa{3h_1+\sqrt{3}h_2}\\
  M_2 &= \Ufactst  \paraa{\sqrt{3}h_1-3h_2}\\
  M_3 &= \Ufactst  \paraa{\ea+\ema+\eb+\emb+\eab+\emab}\\
  M_4 &= \Ufactsti \paraa{\ea-\ema+\eb-\emb-\eab+\emab}\\
  M_5 &= \Ufactst  \parac{\sqo\ea + \frac{1}{\sqo}\ema +
  \eb + \emb + \sqo\eab + \omega\emab}\\
  M_6 &= \Ufactsti \parac{\sqo\ea - \frac{1}{\sqo}\ema +
  \eb - \emb - \sqo\eab + \omega\emab}\\
  M_7 &= \Ufactst  \parac{\frac{1}{\sqo}\ea + \sqo\ema +
  \eb + \emb + \frac{1}{\sqo}\eab + \frac{1}{\omega}\emab}\\
  M_8 &= \Ufactsti \parac{\frac{1}{\sqo}\ea - \sqo\ema +
  \eb - \emb - \frac{1}{\sqo}\eab + \frac{1}{\omega}\emab}
\end{split}\label{eq:MCartan}
\end{align}
%
where \m{\sqo=e^{\frac{2\pi i}{3}}=-\half+\frac{\sqrt{3}}{2}i} (in
this equation, \refeqp{MCartan}).

By considering arbitrary \emph{representations} of \m{\su(3)} one can, also for higher
\m{N} (\m{N\rightarrow\infty}),
obtain a set of matrices,given by \refeqp{MCartan}, satisfying
\refeqp{Mcom}, \refeqp{Asq}, \refeqp{MAcom}.

When checking that \refeqp{Msu} solves \refeqp{Mcom}, one uses that,
(\m{N} arbitrary)
\begin{align}
  \CCom{\UmN}{\UnN}{\Us_{-\nv}^{(N)}} =
  \frac{N^2}{4\pi^2}\sin^2\frac{2\pi}{N}\paraa{\mv\times\nv}\UmN,
  \label{eq:tripleU}
\end{align}
and \m{\sin^2\frac{2\pi}{3}=\sin^2\frac{4\pi}{3}}.

%
%

Similarly, one may take
\begin{align}
\begin{split}
  \Mv = \half\parac{
    \frac{\Um+\Umb}{2},&\frac{\Um-\Umb}{2i},
    \frac{\Ump+\Umpb}{2},\frac{\Ump-\Umpb}{2i},\\
    &\frac{\Un+\Unb}{2},\frac{\Un-\Unb}{2i},
    \frac{\Unp+\Unpb}{2},\frac{\Unp-\Unpb}{2i}
  },
\end{split}\label{eq:MU}
\end{align}
with
\begin{align*}
  \mpv=\begin{pmatrix}
    -m_2 \\ m_1
  \end{pmatrix}\qquad
  \npv=\begin{pmatrix}
    -n_2 \\ n_1
  \end{pmatrix},
\end{align*}
which is a solution of \refeqp{Mcom} for \m{N=\Nh:=\mv^2+\nv^2} (which
we assume to be odd), write the \m{M_j}'s (8 \m{\Nh\times\Nh}
matrices) as (\m{\Nh^2}-dependent) linear combinations of a (\m{\Nh}
``independent'') basis of \m{\glNhC}
\begin{align}
  &M_j^{(\Nh)} = \sum_{a=1}^{\Nh^2-1}\mu_j^a (\Nh)
  T_a^{(\Nh)},\label{eq:MT}\\
  &\Comb{T_a^\Nhp}{T_b^\Nhp} = if_{ab}^c T_c^\Nhp\label{eq:Tcom}
\end{align}
and then define
\begin{align}
  M_j^\Np:=\sum_{a=1}^{\Nh^2-1} \mu_j^a(\Nh)T_a^\Np
\end{align}
to obtain corresponding solutions for \m{N>\Nh} (by letting
\m{T_a^\Np} be \m{N}-dimensional representations of \refeqp{Tcom}).

In the case of \m{\mv^2} being equal to \m{\nv^2}, this detour is not
necessary, and \refeqp{MU} \emph{directly} gives solutions of
\refeqp{Mcom} for \emph{any} (odd) \m{N}. The reason is that, by using
\refeqp{tripleU} the ``discrete Laplace operator''
\begin{align}
  \Delta_\Mv^\Np :=
  \sum_{j=1}^d \CCom{\,\,\cdot\,\,}{M_j}{M_j},\label{eq:dlaplace}
\end{align}
when acting on any of the components of \m{\Mv}, in each case
yields the same scalar factor (``eigenvalue'')
\begin{align}
  \frac{N^2}{4\pi^2}\parac{
    \sin^2\frac{2\pi}{N}\paraa{\mv\times\nv}+
    \sin^2\frac{2\pi}{N}\mv^2+
    \sin^2\frac{2\pi}{N}\paraa{\mv\cdot\nv}
  }.\label{eq:eigen}
\end{align}
The \m{N\rightarrow\infty} limit of this construction gives (a
solution of \refeqp{mpbpb})/\refeqp{rholaplace},
resp. \refeqp{laplace})
\begin{align}
\begin{split}
  \mv(\varphi^1,\varphi^2)=\half\parab{
    \cos\mv\phiv,&\sin\mv\phiv,\cos\mpv\phiv,\sin\mpv\phiv,\\
    &\cos\nv\phiv,\sin\nv\phiv,\cos\npv\phiv,\sin\npv\phiv
  },
\end{split}\label{eq:mcossin}
\end{align}
which for each choice
\begin{align*}
  \mv=\begin{pmatrix}
    m_1 \\ m_2
  \end{pmatrix}
  \quad
  \nv=\begin{pmatrix}
    n_1 \\ n_2
  \end{pmatrix}\quad
  \mpv=\begin{pmatrix}
    -m_2 \\ m_1
  \end{pmatrix}\quad
  \npv=\begin{pmatrix}
    -n_2 \\ n_1
  \end{pmatrix}\quad
  \mv^2=\nv^2
\end{align*}
describes a minimal torus in \m{S^7}.

Interestingly, the \m{N\rightarrow\infty} limit, \refeqp{mcossin},
allows for non-trivial deformations (apart from the arbitrary constant
that can be added to each of the 4 different arguments), namely
\begin{align}
\begin{split}
  \mv_\gamma = \half\parab{
    \cos\gamma&\cos\mv\phiv,\cos\gamma\sin\mv\phiv,
    \cos\gamma\cos\mpv\phiv,\cos\gamma\sin\mpv\phiv,\\
    &\sin\gamma\cos\nv\phiv,\sin\gamma\sin\nv\phiv,
    \sin\gamma\cos\npv\phiv,\sin\gamma\sin\npv\phiv
  }.
\end{split}\label{eq:mgamma}
\end{align}
It is easy to check that \refeqp{mgamma} solves \refeqp{laplace} (and
\refeqp{rholaplace}, with an appropriate choice of \m{\rho},
constant), but when ``checking'' \refeqp{mpbpb} (which is \emph{identical} to
\refeqp{rholaplace}) via the \m{N\rightarrow\infty} limit of
\refeqp{Ucom}, the \m{\gamma}-dependence of the \m{m_j} at first looks
as if leading to a ``contradicition'' (it \emph{would}, in the finite
\m{N}-case), but the rationality of the structure-constants
(\m{\mv\times\nv} instead of
\m{\frac{N}{2\pi}\sin\frac{2\pi}{N}\paraa{\mv\times\nv}})
comes at rescue.

To come to the final observation of this note, rewrite \refeqp{mgamma}
as
\begin{align}
  \mv_\gamma = \frac{1}{\sqrt{2}}\xpg + \frac{1}{\sqrt{2}}\xmg\label{eq:xpm}
\end{align}
with
\begin{align}
  \begin{split}
    \xpg=
    \half\parab{
      &\cos(\mv\phiv+\gamma),\sin(\mv\phiv+\gamma),\cos(\mpv\phiv+\gamma),\sin(\mpv\phiv+\gamma),\\
      &\sin(\nv\phiv+\gamma),-\cos(\nv\phiv+\gamma),\sin(\npv\phiv+\gamma),-\cos(\npv\phiv+\gamma)}\\
    \xmg = \half\parab{
      &\cos(\mv\phiv-\gamma),\sin(\mv\phiv-\gamma),\cos(\mpv\phiv-\gamma),\sin(\mpv\phiv-\gamma),\\
      &-\sin(\nv\phiv-\gamma),\cos(\nv\phiv-\gamma),-\sin(\npv\phiv-\gamma),\cos(\npv\phiv-\gamma)}\\
  \end{split}
\end{align}
While \m{\gamma}, in this form, becomes irrelevant (insofar each of the 4
arguments in \m{\xp:=\xv_+^{[0]}}, as well as those in
\m{\xm:=\xv_-^{[0]}} can have an arbitrary
phase-constant), not only their
sum, \refeqp{xpm}, but (due to the mutual orthogonality of
\m{\xp,\done\xp,\dtwo\xp,\xm,\done\xm} and \m{\dtwo\xm}) \emph{both} \m{\xp} and \m{\xm}
\emph{separately}, in fact any
linear combination
\begin{align}
  \xv_\theta = \cos\theta\xp+\sin\theta\xm\label{eq:xtheta}
\end{align}
gives a minimal torus in \m{S^7}.


\section*{Acknowledgement}

\noindent We would like to thank M.Bordemann, as well as F.Pedit, for
discussions, and S.T.Yau for correspondence.

\bibliographystyle{unsrt}

\begin{thebibliography}{2}

\bibitem{Hoppe:1999}
J.~Hoppe.
\newblock ``Some Classical Solutions of Membrane Matrix Model Equations'',
\newblock hep-th/9702169,
\newblock {\em Proceedings of the Carg\`{e}se Nato Advanced Study
  Institute, May 1997}, Kluwer 1999.

\bibitem{Lawson}
H. B. Lawson, Jr.
\newblock ``Complete minimal surfaces in \m{S^3}'',
\newblock  Ann. of Math. (2) 92 (1970), p 335--374.

\end{thebibliography}

\end{document}